\documentclass[prb,preprint]{revtex4-1} 
\usepackage{amsmath}  % needed for \tfrac, \bmatrix, etc.
\usepackage{amsfonts} % needed for bold Greek, Fraktur, and blackboard bold
\usepackage{graphicx} % needed for figures

\begin{document}

\title{A First-Year Research Experience: The Freshman Project in Physics at Loyola University Chicago}

\author{Jonathan Bougie}
\email{jbougie@luc.edu}
\author{Asim Gangopadhyaya}
\email{agangop@luc.edu}
\author{Sherita Moses}
\email{smoses2@ggc.edu}
\author{Robert D. Polak}
\email{rpolak@luc.edu}
\author{Gordon P. Ramsey}
\email{gramsey@luc.edu}
\affiliation {Department of Physics, Loyola University Chicago, Chicago, IL 60660, U.S.A.}
\author{Weronika Walkosz}
\email{vwalkosz@lakeforest.edu}
\affiliation {Current affiliation: Lake Forest College, Lake Forest, IL, 60045, U.S.A.}

\date{June 21, 2020}
	
\begin{abstract}
Undergraduate research has become an essential mode of engaging and retaining students in physics. At Loyola University Chicago, first-year physics students have been participating in the Freshman Projects program for over twenty years, which has coincided with a period of significant growth for our department. In this paper, we describe how the Freshman Projects program has played an important role in advancing undergraduate research at Loyola and the profound impact it has made on our program. We conclude with suggestions for adoption of similar programs at other institutions.

\end{abstract}
\maketitle

\section{Background and Motivation}	
In recent decades, faculty members at many colleges and universities have worked to involve undergraduates in research in order to provide them with a deeper and more engaged learning experience. As college-bound students and their families visit university campuses, they often ask whether they would be able to participate in meaningful research projects. Faculty members have responded by making their research more accessible to undergraduates, and increasing numbers of universities are now advertising their record in providing such opportunities to their undergraduate students. Federal government programs such as the National Science Foundation-sponsored Research Experience for Undergraduates (REU) and private foundations such as the Howard Hughes Medical Institute have also stepped in by providing resources. \cite{guterman2007} 

Studies have shown that great benefits follow from undergraduate research experiences.  The undergraduate research opportunity is a powerful tool that enhances student engagement, incites curiosity, and emboldens professional identification.\cite{graham2013} The 2003 Strategic Programs for Innovations in Undergraduate Physics (SPIN-UP) report from the National Task Force on Undergraduate Physics (NTFUP) identified undergraduate research, faculty mentorship, and a high level of interaction among faculty and students as key elements of successful departments. \cite{SPIN-UP}
More recently, the final report of the Joint Task Force on Undergraduate Physics Programs (J-TUPP) includes undergraduate research as an important element in case studies of programs ``which have implemented significant activities to prepare their physics students for diverse careers.''\cite{J-TUPP}
Physics graduates report the necessity of skills such as teamwork, technical writing, oral communication, programming, and the application of physics to practical problems and leadership in managing projects; all of which can be developed within the context of research experience. \cite{mcneil2017}
 
While recognition of the importance of undergraduate research experiences has become commonplace, there remain outstanding questions, such as how to best make research accessible to undergraduates, what exactly constitutes undergraduate research, and when in their academic career students should begin research. \cite{Gangopadhyaya2012} For example, while many opportunities such as REU experiences and internships exist mostly for upper-division physics undergraduates, mentoring and student engagement is an important part of attracting and retaining students in their early years in the major, where substantial erosion in the major can happen. \cite{SPIN-UP, J-TUPP} 

In this manuscript, we discuss the Freshman Projects (FP) program in the Physics Department at Loyola University Chicago (LUC), which is designed to engage all majors in the department in research starting from their first year in our program. This program has been running continuously since the 1995-1996 academic year. In the following sections, we describe FP at LUC, discuss its role in the development of our department, discuss the perspectives of various faculty members involved in its implementation, and make some suggestions as to how similar initiatives could be developed at other institutions.

\section{History and Current Status of The Program}

In the Spring 1996 semester, the first cohort of nine freshmen participated in an \textit{ad hoc} research initiative that later gave rise to the FP program. The instructor for the introductory course for physics majors (AG - one of the authors of this manuscript) asked several of his colleagues to serve as mentors to provide engaged and deep learning experiences beyond what could be normally covered in the classroom.

Four groups were formed, and each group of students worked on a project with their respective mentor during that semester. The four projects that students and their mentors chose to investigate were: the Inverse Feynman Sprinkler, \cite{feynman1985} characteristics of high-friction surfaces, a study of a rolling sphere on a curved surface, and the dynamics of motion in vertical circles. 
After designing and performing the experiments, students presented their results to the department.  Students and faculty considered the effort to be a great success.

Given this enthusiastic reception, the department decided to continue these projects with each subsequent cohort of first-year majors. The FP program was initially integrated into the second-semester introductory physics lab for majors (PHYS 126L). In 2008, FP became a separate one-credit course (PHYS 126F). This course is currently required for all students seeking a B.S. in Physics or any of the interdisciplinary majors that are jointly offered by the Physics Department. These interdisciplinary majors include the B.S. in Biophysics, B. S. in Physics with Computer Science, and the B.S. in Theoretical Physics/Applied Mathematics. We henceforth refer to the complete list of majors who are required to take this course as ``majors in the department'' or with  similar nomenclature. 

 Currently, students take FP during the Spring Semester of their first year as a physics major. They generally take this course concurrently with  General Physics II (PHYS 126) and the accompanying Laboratory course (PHYS 126L).
Each participating faculty member is assigned a section of PHYS 126F, with a maximum enrollment of 3-5 students.  
Each group must choose a topic of research in consultation with their mentor. Some faculty mentors choose a project or key area of interest and encourage students who share that interest to work with them, while others develop the project ideas through guided discussion with their students. \cite{departmentwebsite}

Once the Spring Semester begins, the faculty mentor and their students agree on a project to explore and write a simple proposal outlining their plans. Proposals vary in length but are generally at least two or three pages. Each proposal should include a statement of the problem and its motivation, a brief background literature review including key terms and concepts, a list of materials needed to complete the project, and an accompanying budget (current maximum is \$200 per group, paid for by lab fees). In addition, it needs to outline the proposed work itself with an estimated timeline to accomplish the project.

Each group must then implement the proposed project, which must include designing and building an experiment, carrying out related theoretical calculations, and collecting and analyzing data.
As a regular course, each group is scheduled to meet with their mentor once a week for two hours throughout the semester, with additional work performed between meetings. Groups have generally been required to keep a scientific notebook to record their activities throughout the semester.

At the end of the semester, the department sponsors a seminar where each group presents their work, generally as a 12-minute presentation followed by a 3-minute question-and-answer period. The Physics faculty, other students, and guests attend this event and the groups have a chance to answer questions and discuss their work.   Over the years, final reporting has at times incorporated additional written and/or poster presentation components, but the oral seminar presentation has always been an important element. This format gives students the experience of publicly presenting their work, allows students to see the work of the their classmates, and models the format of typical presentations given at professional conferences. We have found it to be a key part of the project.
           
While the projects are appropriate for first-year physics students and do not always involve cutting edge research, the course structure is designed to model several aspects of scientific research: from proposing a project, to carrying out the elements of the proposal, to presenting the results to a wider audience. Since these projects are open-ended and are not to be found in any handbook, there is always the ``thrill'' of discovery in the air.

\section{Benefits to Students and the Department}

\subsection{Project-based Learning}

The benefits of research engagement are vast both for students and research mentors, impacting not only their cognitive and intellectual growth but also leading to professional advancement. \cite{osborn2009}  While research engagement for undergraduates is generally optional and typically involves upper-level students, FP at Loyola is mandatory and involves students in their first year of college.

This experience integrates various skill sets in investigating an applied problem just beyond the level of the freshman courses. Learning outcomes for the course include deepening students' understanding of introductory physics concepts and familiarizing students with research methods employed in the field. 

Whether the topic is chosen by the mentor or suggested by a student, guiding the group in choosing a suitable topic is an important part of the faculty mentor's role and can help ensure a successful learning experience for students. The topic should be challenging enough that it will keep students engaged for a semester while also being sufficiently tractable that first-year students can make significant progress and make a meaningful presentation. One possible guiding principle is to make projects somewhat scalable, with goals of varying difficulty levels. That way, if unexpected challenges arrive, some progress will still have been made, while there is room to expand the project if everything goes smoothly, possibly even beyond the FP. 
			
One inherent part of research is dealing with unexpected results, and projects rarely work perfectly as intended at the first try. When a design initially fails or when experimental results do not match theory, students learn to troubleshoot and check their assumptions.   We have learned that having a sense of ownership of a project helps to actively engage students, and they frequently challenge themselves and take significant initiative in developing and furthering their projects. Faculty judgment is important in determining when to step back and let students wrestle with a problem, when to teach them a new concept or method, and when to guide them in a new direction. Students learn about scientific ethics; if attempts to fix problems fail, data should not be altered to fit results. Instead, the end result of the research may be to find explanations for the failure and to learn lessons for the future.

These projects help students develop a variety of skills. Depending on the specific project, these can include various experimental techniques and the use of the machine shop. Students may learn programming skills or mathematical methods that may exceed what they would normally learn in their first year. Learning these skills in conjunction with a specific project emphasizes the application of this material.
	
Regardless of the specifics, the experience of approaching an open-ended problem and developing the knowledge and tools to tackle that problem prepares students for project-based learning in their advanced laboratory classes. Faculty in upper-level courses can expect that our majors know how to delve into problems in more depth than end-of-chapter problems they find in their textbook and how to explore a question when the answer may not be known. This experience can also help students develop an interest in pursuing future research opportunities.

By working together as a group during the semester, students learn important teamwork and project management skills and can form close bonds with their teammates and mentor. Such bonds have been important in integrating students into our department; helping to attract and retain students. By writing project proposals and presenting their final results, students learn how to present scientific results both in oral and written forms. 	 In these ways, FP fulfills many of the goals established by the	J-TUPP report for preparing students for the contemporary job market. \cite{J-TUPP} 

Specific assessment rubrics have varied over time and by faculty member. In line with the course goals, students are expected to participate actively throughout the semester, work collaboratively as a part of the team, propose and carry out a well-designed investigation into a chosen topic, and to clearly and professionally present their work to the class.  With appropriate guidance, engaged students can deepen their understanding of physics and learn important research methods and approaches regardless of whether all goes as expected and they confirm their initial hypothesis or not. Therefore, our consistent experience is that most groups successfully achieve the desired outcomes of the course.

With the longevity of our program, we now have a former student who returned to Loyola as a faculty member (WW- one of the authors of this manuscript) and who is now at Lake Forest College. She has now seen the transformative aspect of the FP from the perspective of a mentor. Here WW describes her experience:

{\bf VIGNETTE: Dr. Weronika Walkosz}

 As someone who participated in FP as both a student and faculty member, I would like to outline some of the advantages of Loyola's FP program on learning, attitude, and professional development of freshman students and their mentors.  
 
First, my early involvement in research as a freshman student helped me become easily integrated into Loyola's Physics community.   By working closely with other freshmen and a mentor on a common project, I was able to form professional relationships that helped me navigate a demanding physics curriculum.   Indeed, my FP partner and I continued learning from each other, preparing for exams and solving homework problems together.  The project also provided me with an ongoing one-on-one mentorship with a physics faculty member who could advise me better on my career options.  The project also helped me develop good communication and presentation skills and made me more confident to ask questions in class.  

Second, the early exposure to research taught me personal responsibility and persistence in solving difficult and often unexplored problems, prepared me for advanced coursework, and trained me on how deal with uncertainty.  I learned to balance my course work with my research activities, collaboration with independence, factual knowledge with freedom of thought and discovery.   The project helped me realize that even simple phenomena have many nuances, and their exploration and discovery is not always easy or straightforward.  

Third, the project was instrumental in clarifying and preparing me for a career in science.  Through my participation in FP, I have learned how to think, plan, and design as a scientist.  After completing the FP in my first year of college I was able to continue the work in the following year and apply for scholarships to fund it.   A longer commitment to my research project, in turn, prepared me better for graduate school.  The possibility of continuing the work after the freshman year is one of the most important features of the FP at Loyola that can lead to peer-reviewed publications, scholarships, participation in national conferences, or simply help in post-graduation plans. 
 
As a faculty member supervising a project, I was able to see my students participate in the benefits of project-based learning in a way that paralleled my own experience as a student.  The project allowed the students to focus on an open-ended question understanding and solving of which required exploration, analysis, creativity, communication and collaboration.  It challenged them to take ownership of their own learning, while inspiring and motivating each other.  It also showed them that research requires patience, persistence, and good management skills.

\subsection{Ongoing Research and Dissemination}\label{sec-ongoing}

The seminar concluding the semester gives students an experience developing presentation skills and communicating scientific ideas. In addition, many projects initiated as FP are presented at the local chapter of the American Association of Physics Teachers (AAPT) or other symposia. In 2005, our projects were featured in the \textit{Chicago Tribune}. \cite{chitribune}

Some FP have blossomed into more advanced research projects and have been presented at regional, national, and international conferences.  Some have also resulted in publications with students as co-authors, particularly in journals such as \textit{Physics Teacher} and \textit{American Journal of Physics}.  

To give a sense of the breadth of topics covered in Freshman Projects, as well as the potential for some projects to develop beyond the initial investigation, we list here some recent publications that began as FP, each with student authors or co-authors: 

\begin{enumerate}
\item A series of FP groups has worked with author GR to study properties of musical instruments such as instrument shape, size, and method of excitation, and to correlate these to acoustical properties using tools such as frequency analysis and high-speed photography. These projects have led to a series of presentations and publications.\cite{ramsey2012, wiseman2012, pomian2014, ramseypomian2014}
\item A group working with author AG conducted a study of magnetic damping by using a two-pulley system to conduct a controlled drop of a neodymium magnet through a copper pipe. By using smart pulleys, they recorded the position, velocity, and acceleration of the magnet as it passed through the pipe and found its terminal velocity as determined by magnetic damping. Following the FP, they were able to conduct an analytical study resulting in an article in \textit{American Journal of Physics}. \cite{irvine2014}
\item Author RP has conducted an ongoing series of FP with students regarding experiments in optics, discussed in more detail in the vignette later in this section. \cite{polak2014, polak2016, polak2017, polak2018} He also worked with a group to use an iPhone app to measure sound frequencies produced by a guitar string; they tested Young's modulus by measuring the change in string length created by the gearing of the tuning pegs. \cite{polak2018b} 
\item Author SM worked with a group of students who were interested in her field of nanobiophotonics. She worked with her students to design a project to give them experience using research techniques and instruments used in this field to reduce and study gold nanoparticles and their effect on \textit{Escherichia coli} bacteria. \cite{moses2018} This project is discussed further in the vignette in Sec.~\ref{sec-smoses}.
	
\end{enumerate}

It is important to recognize that many students who become excited about their research in the freshman year continue research through later years, and not always in the same area as the first year. Therefore, the FP program has encouraged a culture of undergraduate research for students and faculty in our department that goes beyond publications directly related to FP.

As one example, several groups of students have worked under the supervision of RP (a faculty member and one of the authors of this manuscript) on a series of optics projects with the idea of developing low-cost optics experiments and demonstrations with classroom application. These groups have presented their results at various professional meetings and have led to several journal publications. Below, RP describes how his work with students got started:

{\bf VIGNETTE: Dr. Robert Polak}

We initially developed a project to create a low-cost student spectrometer and a demonstration of image formation using LEDs and a cylindrical lens. We built the spectrometer using PVC tubing,  dowels, a black drape, a protractor, and a diffraction grating. With these low-cost materials, we were able to build a spectrometer that could measure the angular location of the zeroth and first-order fringes of a colored LED and used it to calculate the wavelength of that LED, yielding a result within 5 nm of the accepted value. We also built a low-cost light source using LEDs, resistors, and a 9-V battery. We used a white panel board to follow the path of the light and to mark locations of the object, lens and image.  This setup can be used to measure magnification and confirm both the lensmaker and thin-lens equation.

 We presented this project at a local AAPT meeting where it was suggested that we should attempt to publish it.  Working with the students over the summer, we completed a manuscript that was accepted for publication in \textit{The Physics Teacher}.\cite{polak2014}  We developed more experiments and demonstrations over the next two years, starting with understanding easily observable properties of waves using ripple tanks such as interference and diffraction and then showing how to demonstrate interference and diffraction of sound and light using readily available materials. \cite{polak2016}  This has continued to be a source of inspiration for FP as the physics is easily accessible for first-year Physics Majors and the development of low-cost instructional tools meets social justice goals of Loyola University Chicago.

Individual students involved in these projects have often found a path in their academic career.  In one case, a transfer student demonstrated particularly strong experimental skills in creating these new experiments.  She continued to work on the project for the next year, guiding some work to publication.  Based on her strong experimental skills, she earned the opportunity to participate in an REU experience and then completed her senior year working with an experimental scientist at Argonne National Laboratory, providing her the skills needed to pursue graduate studies in optics.

\subsection{Integration of Faculty Into the Department}\label{sec-smoses}

In addition to its benefits to students,  FP can have benefits for faculty in the department. As well as helping faculty members to find students to work with in their research, working closely with students and developing collaborative projects has important benefits in integrating new faculty into the department.

FP can initiate collaboration between faculty members as well, giving faculty opportunities to discuss mentoring strategies and share experiences working with students. Before they mentor their first project, new faculty should have had the opportunity to hear about some of the previous projects and to ask questions of other faculty members. Faculty can be encouraged to discuss their projects with each other, to share resources, and to help each other when needed. Department meetings can be used to discuss preparation for the project and to check-in on its progress, and FP can be useful in framing discussions of pedagogy and mentoring. All faculty in the department should be invited to the seminar at the end of the Spring semester. Therefore, even if a faculty member is not mentoring a project in a given year, they can be included in a shared experience making the FP a part of the life of the department. 
	
As one example of the benefits of FP for integrating faculty into the department, SM (one of the authors of this manuscript) discusses her experience teaching FP in her second year in the department:

{\bf VIGNETTE: Dr. Sherita Moses}

Students chose to enroll in my section based on their interest in my area of research and submitted an interdisciplinary proposal involving nanobiophotonics. The initial class discussed what would be required to complete the research, including the collaborative interdisciplinary effort that would be required. My goal was to create hands-on experiences by exposing my group to different laboratories to train on state-of-the-art instruments of science. This provided me the opportunity to interact with other scientists in the academic, as well as the professional community. The resulting research was published,\cite{moses2018} and those undergraduate students are now active members in the Research Gate community.

In my opinion, the benefits of teaching FP included increased student involvement in my area of research and building student-faculty interactions. This teaching experience resulted in an addition to my publication portfolio. Finally, what I found to be particularly rewarding were the new relationships formed with other academics and scientists across different disciplines, forging relationships that will foster future opportunities for original research. FP is an ideal course to use to develop your faculty members on all tiers. Understanding your vision, believing in life-long growth, establishing a great network, taking the initiative, and at all times being a person of integrity exhibit some of the fundamentals of faculty development. \cite{phelps2016} Participating in this project at Loyola University Chicago provided these fundamentals for me.

\subsection{Impact of  FP on a growing department}

The project has had a very positive impact on our program since its inception. It has played a key role in developing and cohering a rapidly growing department as it has coincided with a period of strong growth. As Table \ref{table-graduationrate} shows, our department has had a great success in attracting and retaining growing numbers of students since the introduction of FP. The graduating class of 1998 was the first class that participated in FP, and there has been a large increase in the number of graduates since then. 
\pagebreak

\begin{table}[h!]
	\centering
	\caption{Number of graduates from the LUC Physics Department as averaged over seven separate four-year periods. \cite{graduationrecords} Each year listed indicates the academic year ending in that year; for example, ``1991'' indicates AY 1990-1991.}
	\begin{ruledtabular}
		\begin{tabular}{l p{5cm}}
			Year range  & Average number of grads per academic year\\
			\hline
			1991-1994 & 5.0 \\
			1995-1998 & 5.3 \\
			1999-2002 & 7.5 \\
			2003-2006 & 9.0 \\
			2007-2010 & 23.8\\
			2011-2014 & 27.8 \\
			2015-2018 & 30.3\\
		\end{tabular}
	\end{ruledtabular}
	\label{table-graduationrate}
\end{table}

This project was an important part of a series of curricular changes that has led to the revitalization of the department. It has played a role in building a familial atmosphere in a department with students occupying a central place in the unit at an important stage in their careers. As the department grew, attention was needed to keep faculty members accessible to students. Otherwise, students who once could expect relatively individualized attention from their freshman sequence instructors could easily get lost. Through their FP mentor, students have another go-to-person to discuss their concerns or to just chat about their future. They get the opportunity to work closely with a mentor and their peers, and these experiences can help them to identify closely with the department and to feel welcomed and integrated into the activity of the department.

\section{Suggestions for Development of Similar Programs}

\subsection{Faculty Involvement}

In the FP course, students and faculty members spend a large amount of time working together to complete their projects within a semester. Faculty mentoring plays an important role in guiding students through the project. To ensure success, chosen projects should be sufficiently complex yet not overwhelming, with well-established goals that can and should be met through collaboration.   The role of a mentor is crucial in helping students individually understand the problem and work together to solve it. Although all students should be actively involved in each aspect of the project, facilitating different students to lead different aspects of the project gives students a chance to develop collaborative leadership skills. Therefore, the most important key to success of the program is the buy-in by faculty members. 

The benefits of the program to the faculty and to the department can help ensure faculty commitment. Faculty can learn more about the physics major at an early stage in their departmental involvement. Students can be attracted to the department by seeing the exciting activity of other students and introducing students to research early in their careers can increase student retention. The potential benefits for the department are therefore significant. 

At Loyola, there was initially no formal credit given to faculty mentors,  and yet faculty involvement was strong due to their appreciation for the potential benefits of the course for the students and the department. This, however, was not an ideal situation as faculty took on additional responsibilities for the project without receiving credit; equally importantly, the transcript of the students did not show that they had carried out research with their advisors during their freshman year. Therefore, additional incentive can be provided for faculty if their effort on the project can be counted in some manner towards their teaching responsibilities, and sometimes to their scholarship. At Loyola, we did this in 2008 by making FP a separate one-credit course.

\subsection{Physical Resources}

The physical resources required to start a similar program are small. Minimum necessary resources include: work space (one table per group) and access to lab equipment and computing facilities (which may be as simple as standard laptop computers).   While material supplies needed for building many experimental setups can be acquired at home improvement stores at a reasonable cost, some additional funds may be required. We started the projects at Loyola with an average budget of \$50 per group for supplies, which has been gradually increased to its current rate of \$200 per group. At Loyola, we have been able to provide this funding by instituting lab fees for lab-based courses in Physics, including FP. The collection of lab fees has occasionally also allowed us to purchase equipment that can be shared in various FP as well as advanced undergraduate projects.  Other resources available to the department can also be usefully integrated into the FP course. As an example, we have a machine shop at Loyola that is staffed by a skilled part-time employee. This machinist provides a safety class to FP students and trains them in the use of several tools.
The benefits in attracting and retaining students may also justify the allocation of additional institutional resources to the department.

\subsection{Integration Into Curriculum}

When considering such a program, it is important to assess its desired role in the curriculum and how to introduce this new element successfully.
One possibility is to start it as a part of the second semester of the introductory physics lecture or the corresponding lab course, as we did at Loyola. This allows the program to get off the ground with a relatively minor commitment. Assigning a faculty FP coordinator (which could be the instructor of the introductory physics class or lab) can help to unify activity across the various groups. If possible, it may be helpful to start the project as an innovative course, i.e., develop a set of learning objectives,  specific goals, time commitments, grading rubrics and responsibilities for students and faculty members before implementing the program, as well as an appropriate assessment plan for future improvements. 
 
It is generally useful to consider the FP experience as a simple model of physics research, beginning with a proposal that includes requirements for theoretical, experimental and construction components. As the project progresses, students should document their progress electronically and/or in a lab notebook. Finally, a seminar or poster session to complete the semester gives students experience in scientific presentation.

To maximize the impact of the activity, groups can be encouraged to present their work outside the department, such as at a local AAPT meeting, an undergraduate research symposium or a Society of Physics Students regional or national meeting. Depending on the nature of the project, faculty members and their research groups could consider publication in an undergraduate research journal, or journals such as {\it The Physics Teacher}, {\it American Journal of Physics}, or journals devoted to research on specific areas.

The benefits of the program may trickle in slowly, hence some level of patience is warranted.
The faculty should meet as a whole to periodically assess progress and discuss the projects based on experience and student feedback. There should be a review of the projects at the end of the semester to decide on any necessary changes for future years. As a part of their formative first year at LUC that informed their overall experience in our department, we have gotten significant feedback from students regarding FP through Senior exit interviews as they prepare for graduation and reflect on their experience at Loyola.

Naturally, implementation of these suggestions may need to be modified based on the available faculty members in a department, its student body, and the resources available. However, even a subset of these elements, combined with a department's ideas for their own project development, could provide an excellent FP experience for a cohort of physics majors.  

\section{Conclusions}	

The Freshman Project in Physics has played an important role at Loyola University Chicago as the department has transformed from a relatively small major to a department with a number of annual graduates frequently among the top ten undergraduate-only Physics programs in the country.\cite{mulvey2012} In doing so, it has helped to facilitate interaction between faculty and students, and has taught students valuable skills that will prepare them for careers in the Twenty-First Century. We believe that this program can provide a useful model for incorporating project-based learning at early stages in students' careers into the curriculum at other institutions. 

\section{Acknowledgments}

The authors would like to thank all of the LUC Physics faculty who have contributed to the development of the FP program. We would also like to thank Dr. Maria Udo for discussion during the preliminary stages of this manuscript, and Mr. Thomas Ruubel, whose excellent record keeping provided some data related to Table ~\ref{table-graduationrate}. We also thank the anonymous reviewers, whose useful questions and suggestions contributed to the revision process.

\end{document}